\documentclass[aps,prl,twocolumn]{revtex4}

\newcommand{\vn}{{\vec n}}

\begin{document}


\title{Comment on "Critical Dynamics of a Vortex-Loop Model for the
Superconducting Transition"}

\author{Jack Lidmar}

\affiliation{Department of Physics,
Stockholm University, SCFAB,
SE-106 91 Stockholm, Sweden}

\date{\today}



{\noindent \bf Comment on ``Critical Dynamics of a Vortex-Loop Model
for the Superconducting Transition''}

\medskip

In a recent Letter~\cite{AG}, Aji and Goldenfeld (AG) study the
critical dynamics of the normal- to superconducting phase transition
in zero magnetic field.  They study both continuum models of vortex
loops and discrete lattice models, often used in numerical
simulations, and come to the surprising conclusion that their dynamic
critical behavior differ.  In this Comment we point out a serious
problem in their analysis of the lattice models.

The critical behavior of type-II superconductors can be characterized
by large scale fluctuations of vortex loops.  Close to the phase
transition, it is reasonable to assume overdamped relaxational
dynamics for the vortex degrees of freedom, such as generated by a
Monte Carlo (MC) simulation.  Simulation studies of lattice models
have shown that the resulting dynamic critical exponent $z$ is given
by $z \approx 1.5$ for unscreened vortex interactions and $z \approx
2.7$ if the vortex interaction is assumed to be strongly
screened~\cite{WJ,LWWGY}.  In high temperature superconductors, where
critical fluctuations are most pronounced, the screening length is
very large but finite.  Under renormalization the screening length
shrinks and hence both these limits are interesting to consider.  The
values of the critical exponents are surprising since the naive
expectation from relaxational (Model-A) dynamics is an exponent close
to $z\approx 2$~\cite{HH}.

In their Letter, AG propose an explanation for the simulation results,
but also argue that the lattice models do not capture the true
continuum behavior which should instead have a dynamic exponent $z
\approx 2$.  They claim that the discrepancy is due to an incorrect
identification of Monte Carlo time with real time.  If true, this
could have important consequences for many other studies of dynamics
where one usually assumes that the artificial discreteness from a
lattice is irrelevant for the critical properties, and that $t_{\rm
real} \sim t_{\rm MC}$.  Therefore this topic deserves careful study.

AG base their scaling analysis of the vortex Hamiltonian on the
assumption that the coupling constant $J$, which is dimensionless in
the lattice model, is not being renormalized, and hence does not carry
any scale dependence.  Requiring extensivity of the free energy they
then arrive at a scaling dimension $x$ of the vorticity $\vn \sim
L^{-x}$, which is $5/2$ in case of long range unscreened interactions
and $3/2$ for short range interactions.  However, the only reason why
the coupling constant $J$ is dimensionless in the discretized model is
because it is multiplied by the appropriate power of the microscopic
lattice spacing $a$.  Therefore this does {\em not} exclude any
nontrivial scale dependence of $J$.  In fact, more standard scaling
arguments give (see e.g.\ Ref.~\cite{LW-BG}) that $\vn \sim L^{-2}$,
i.e., $x=2$.  This follows because $\vn = \nabla \times \nabla
\theta/2\pi$, where $\theta$ is the phase of the superconducting order
parameter.  $\theta$ in turn has to have the trivial scaling dimension
$0$, since it should be $2\pi$-periodic at any scale.  Only with this
scaling follows the well known finite size scaling relations for the
superfluid density, $\rho_s \sim L^{2-d}$, and for the magnetic
permeability, $\mu \sim L^{d-4}$, where $d$ is the dimension.  These
are purely static quantities, whose scaling can be easily checked in
numerical simulations without any assumption about the proper relation
between real and MC time.  This has been done many times for the
models under consideration and is in fact one of the best ways to
locate the transition~\cite{LWWGY}.  The scaling dimensions proposed
by AG would instead lead to the predictions $\rho_s \sim L^{2x-d-2}$,
$\mu \sim L^{d-2x}$, which are not supported by MC data.

In conclusion, the explanation of the simulation results offered by AG
is not acceptable, since it is based on incorrect assumptions.  There
is no need for a scale-dependent correction in the proportionality
between real time and MC time, and hence no correction to the dynamic
exponents found in simulations.  It remains to explain why these
values are so far from the naive expectation.

Discussions with Nigel Goldenfeld and Mats Wallin are gratefully
acknowledged.  This work was supported by the Swedish Foundation for
International Cooperation in Research and Higher Education (STINT) and
the Swedish Research Council (VR) through Contract No.\
F~5104-278/2001.

\bigskip

\noindent
Jack Lidmar

\medskip

\noindent
{\small
Department of Physics\\
Stockholm University, SCFAB\\
SE-106 91 Stockholm, Sweden
}

\medskip

\noindent
Received 4 December 2001;
Published 15 August 2002 in \prl {\bf 89}, 109701 (2002).

\end{document}